**Comment on "Design and circuit simulation of nanoscale vacuum channel transistors"**

Richard G. Forbes[a]

Abstract

These comments aim to correct some apparent weaknesses in the theory of field electron emission given in a recent paper about nanoscale vacuum channel transistors, and to improve the presentation of this theory. In particular, it is argued that a "simplified" formula stated in the paper should not be used, because this formula is known to under-predict emission current densities by a large factor (typically around 300 for an emitting surface with local work function 4.5 eV). Thus, the "simplified" formula may significantly under-predict the practical performance of a nanoscale vacuum channel transistor.

[a]*University of Surrey, Advanced Technology Institute & Dept. of Electrical and Electronic Engineering, Guildford, Surrey, GU2 7XH, UK. E-mail: r.forbes@trinity.cantab.net*

In a recent paper[1], Ji Xu, Yaling Qin, et al. have discussed the design and circuit simulation of nanoscale vacuum channel transistors, assuming that the electron emission mechanism is field electron emission (FE). By modern standards there are some weaknesses in their presentation of FE theory, and a "simplification" that they have made would lead to significant under-prediction of emission current densities and emission currents. These comments aim to correct weaknesses and set out a clearer presentation of relevant FE theory. The nanoscale vacuum channel transistor is potentially an interesting technological device, and for research and development purposes it is important that theoretical analyses of its behaviour should conform to best current practice.

Ji Xu et al. make the usual (acceptable) simplifying assumption that their emitter can be modelled theoretically as if it were a free-electron metal with a smooth planar surface of large lateral extent, and start from what is in fact the zero-temperature form of the FE equation developed by Murphy and Good[2] in 1956. (For a modern derivation, see Ref. 3.) The Murphy-Good FE equation corrected errors found in the 1950s in the original 1928 treatment by Fowler and Nordheim.

Ji Xu et al. start by writing the Murphy-Good FE equation for local emission current density $J$ in the detailed form

$$J = \frac{e^3 F^2}{8\pi h \phi t^2(y)} \exp\left[-v(y) \frac{8\pi (2m)^{1/2} \phi^{3/2}}{3heF}\right], \qquad (1)$$

where $e$ is the elementary (positive) charge, $m$ is the electron mass, $h$ is Planck's constant, $\phi$ is the relevant local work function, and $F$ is the relevant local surface field. The Nordheim parameter $y$ is a known function of $\phi$ and $F$, and $v(y)$ and $t(y)$ are appropriate particular values of well-defined special mathematical functions (e.g., see Jensen[4]). Note that, because—in the modern International System of Quantities[5]—eq. (1) is dimensionally consistent, it is not necessary to specify the units in which the quantities in eq. (1) are measured.

Modern discussions of FE theory normally simplify eq. (1) by defining two universal constants, denoted here by $A$ and $B$ and often called the first ($A$) and second ($B$) Fowler-Nordheim (FN) constants. The definitions and values (to 7 significant figures) of these universal constants can be written

$$A \equiv e^3/8\pi h \cong 1.541434 \times 10^{-6} \text{ A eV V}^{-2} , \tag{2}$$

$$B \equiv 8\pi(2m)^{1/2}/3eh \cong 6.830890 \times 10^{9} \text{ eV}^{-3/2} \text{ V m}^{-1} . \tag{3}$$

The constants are known to at least the accuracy stated, but should be suitably rounded in technological contexts. Inserting these constants into eq. (1) yields a common modern form for the Murphy-Good FE equation:

$$J = \frac{AF^2}{\phi t^2(y)} \exp\left[-v(y)\frac{B\phi^{3/2}}{F}\right] . \tag{4}$$

Note that, in the modern International System of Quantities introduced in the 1970s, $A$ and $B$ are not pure numbers but are physical constants with units. Failure to state the units can lead to dimensional inconsistency and discrepancies. Thus, just below their eq. (2), Ji Xu et al. give the value of $B$ as $6.83 \times 10^7$: this numerical value has probably been taken from a paper where fields were being measured in V/cm. However, in their eq. (5) the value $6.83 \times 10^9$ is used, which is the appropriate numerical value when fields are measured in V/m.

In fact, eqns (4) and (5) in the paper under discussion are dimensionally inconsistent, because the units of $A$ and $B$ have been omitted. In the modern ISQ, the best way of writing these equations is to put the FN constants $A$ and $B$ as defined above into these equations, rather than their numerical values, and give the values of $A$ and $B$ separately, as above.

Ji Xu et al. next make two modifications to eq. (4) above. In the first of these, they attempt to relate the *local* emitter surface field $F$ to (what appears to be) the average field between the emitter and a counter-electrode. This average field is denoted here by $F_{av}$, and may be significantly smaller

than the local surface field $F$, with the relationship between the values of the two types of field given by

$$F = \beta F_{av}, \tag{5}$$

where $\beta$ is a field enhancement factor. In the paper under discussion, this relationship is confused by the fact that the authors use the same symbol "$F$" for both of the two different types of field. However, provided that numerical methods have been used to calculate the local field, and that a field emission formula involving *local* field has been used to calculate local current density, no problem should arise from this confusion over field definitions.

The authors then attempt to formulate a relationship between average field $F_{av}$ and the voltage $V$ between the emitter and the counter-electrode. The formula that they require is actually

$$F_{av} = V/d, \tag{6}$$

where $d$ is the distance between the emitter and the counter-electrode. However, in the text below their eq. (2) they give the relation as: $F = V \times d$. This formula cannot be correct, because it gives the units of field as V m, rather than V/m. As a consequence, the authors' formula (5) is also incorrect. In the approximation that the authors are using, their eq. (5) should be written

$$b = \frac{B\phi^{3/2}d}{\beta}. \tag{7}$$

where the meaning of $b$ is defined by their eq. (3). Their eq. (4), however, does correctly place $d^2$ in the denominator.

We now return to discussion of my eq. (4) above. The second modification made by the authors is to "simplify" eq. (4) by setting $t(y)$ and $v(y)$ equal to unity, yielding (before conversion to voltage-based form)

$$J = \frac{AF^2}{\phi} \exp\left[-\frac{B\phi^{3/2}}{F}\right]. \tag{8}$$

This equation (which is sometimes called the "elementary FE equation") is undoubtedly simpler, but this approximation also introduces numerical error, namely under-prediction of current density by a large factor[6]. The size of the discrepancy depends markedly on the values of the local work function $\phi$ and the local field. Typical values for the size of the discrepancy range from roughly 400 for a

$\phi$=4.0 eV emitter, to roughly 300 for a $\phi$=4.5 eV emitter, to roughly 250 for a $\phi$=5.0 eV emitter. Ji Xu et al. do not point out that *significant under-prediction* of the performance of the nanoscale vacuum channel transistor is involved in this "simplification".

The view of the present author is that this under-prediction is an undesirable feature of an unnecessary "simplification". Although setting $t(y)=1$ is an acceptable approximation, it is much better to leave the correction factor $v(y)$ in the FE equation. Conversion to current-voltage-based form then yields the equations

$$I = \frac{\alpha A \beta^2 V^2}{\phi d^2} \exp\left[-v(y)\frac{B\phi^{3/2}d}{\beta V}\right], \tag{9}$$

$$\ln\{I/V^2\} = \ln\left\{\frac{\alpha A \beta^2}{\phi d^2}\right\} - v(y)\frac{B\phi^{3/2}d}{\beta V}. \tag{10}$$

where $I$ is the emission current and $\alpha$ is the formal emission area.

An *ideal* experimental FE device/system is one where there are no "complications", such as those associated with series resistance in the current path, or with current dependence in field enhancement factors. For an ideal system the measured current-voltage characteristics can validly be identified with the theoretical emission characteristics. Equation (10) thus describes the form theoretically predicted from Murphy-Good FE theory for the shape of a measured FN plot taken from an ideal experimental system. Because $v(y)$ is an indirect function of $V$, it can be shown that the slope $S$ of the theoretical FN plot is given by

$$S = d[\ln\{I/V^2\}]/d(1/V) = -s(y)\frac{B\phi^{3/2}d}{\beta}, \tag{11}$$

where the slope correction function[7] $s(y)$ is a slowly varying function of $y$ (and hence of $V$). The function $s(y)$ can be adequately approximated as having the constant value 0.95. For an ideal system, the other parameters in eq. (11) are also constants. Thus, an experimental FN plot taken from an ideal FE system is predicted to be "nearly straight". Observed apparent linearity of a FN plot is an indicator that the emission process involved is likely to be field electron emission, as the authors point out, and that the system is behaving in an ideal fashion. There also exists a test[8] that can determine experimentally whether a FE system is ideal.

The following point also needs to be made. Strictly, the theory given above (which relates the emitter local surface field to a measured voltage) is theory that applies to a diode configuration in which voltage $V$ is unambiguously defined. However, the transistors under discussion have three electrodes. Care is needed in order to apply the above voltage-based theory only in circumstances where the transistor is effectively behaving as a diode. In other circumstances, eq. (4) above remains valid, but the distribution of local surface field $F$ may need to be determined by numerical methods.

The remarks in this comment use an approach and notation that aims to be close to that used in the paper under discussion. Usually, the present author's preference would be to formulate FE theory in a somewhat different manner, using different notation conventions described elsewhere[9].

A general consequence of these remarks is as follows. It is not clear from the paper precisely which equations have been used to carry out the reported simulations of circuit behaviour. If equations (2)-(5) in the paper have been used, rather than eq. (1) in the paper, then the reported results will probably be quantitatively unreliable.